\documentstyle[aps,manuscript]{revtex}

\begin{document}

\title{RELAXATION OF FAST COLLECTIVE MOTION IN HEATED NUCLEI}

\author{V.A. Plujko (Plyuyko)
\address{Institute for Nuclear Research, Kiev, Ukraine}
}

\maketitle


\begin{abstract}
The damping of the collective vibrations in hot nuclei is studied 
within the semiclassical Vlasov-Landau kinetic theory. The extention 
of the method of independent sources of dissipation is used to 
allow for irreversible energy transfer by chaos weighted wall formula.
The  expressions for the intrinsic width of the giant multipole 
resonances are obtained. The interplay between the one-body and 
the two-body channels which contribute to the formation of the 
intrinsic width  in nuclei is discussed.
\end{abstract}

\vskip 1 cm

PACS numbers: 21.60.-n,21.60.Ev,24.30.Cz

\section{Introduction}
  The relaxation mechanisms of collective motion and their dependence 
on the temperature  in many-body systems have been much investigated
in recent years \cite{g92}- \cite{bctbb}. In the present paper we consider 
the damping of the nuclear multipole vibrations within the semiclassical 
Vlasov- Landau kinetic theory. Semiclassical methods seem to be quite 
instructive for an investigation of the averaged properties of the 
multiparticle systems. In many cases, they allow us to obtain analytical
results and represent them in a transparent way. 

 In what follows we concentrate on the investigation of the contributions 
of different relaxation mechanisms to the intrinsic width of the giant 
multipole resonance (GMR). We determine the intrinsic width as formed by 
two main sources: {\it 1}) The relaxation due to the coupling of both 
particle and hole to more complicated states lying at the same excitation
energy. This is the so-called two-body collisional damping. 
We take into account the two-body (collisional) damping exactly, 
incorporating into the collision integral the memory effects associated 
with the mean-field vibrations \cite{kmp}- \cite{kp1}; 
{\it 2})~The fragmentation width caused 
by the interaction of particles with the time-dependent self-consistent 
mean field. It was shown in Refs. \cite{yan1,yan2}, in the classical 
limit for the random phase approximation, the fragmentation width 
coincides with the width obtained from the one-body 
("wall"~\cite{my,bl1}~) relaxation 
mechanism. We will imitate the fragmentation width by the one-body 
relaxation. This relaxation  is taken into consideration approximately by 
adding to the Vlasov-Landau equation some source terms. The extention of 
the method \cite{kps} of independent sources of dissipation is used to 
allow for irreversible energy transfer by chaos weighted wall 
formula \cite{ts}. We also do not use the normalization of the width
to  a magnitude corresponding to the infinite- matter value. 

In Sect. 2 the intrinsic width of the giant multipole 
resonances in cold and hot nuclei is calculated. 

The numerical results and general discussion of the mass number 
and temperature dependences of the different relaxation 
mechanisms are presented in Sect. 3. 

\section{ Damping of the giant resonances}

 We will consider a nucleus as  the nuclear Fermi-liquid drop.
We adopt the sharp surface for protons $R_{p}(t)$ and neutrons $R_{n}(t)$
and describe  the edge of the nucleus in terms of the effective surface 
$R =(R_{p}+ R_{n})/2$. The GMR of multipolarity $\lambda $ is regarded 
(see, \cite{kmp,den}) as oscillations of the corresponding (isoscalar or 
isovector) density vibration inside  the nucleus, associated with the 
vibrations of the local displacement of the nucleons 
$
\delta R^{(\pm)}_{\lambda}(t) \equiv 
\delta R_{\lambda,p}(t) \pm \delta R_{\lambda,n}(t) 
\equiv  R_{0} 
\beta^{(\pm)}_{\lambda } (t) Y_{\lambda 0}(\hat{r})
$ 
of the same multipolarity $\lambda $; the signs (+) and (-) stand for 
isoscalar and isovector GMR respectively. Here $R_{0}$ is radius in 
equilibrium and $\delta R_{\lambda,\alpha}$ is the local displacement 
of the particle of the $\alpha = (p,n)$ type from its equilibrium 
position.    

 The density  vibrations of the every kind (isoscalar or isovector) 
is defined by the variation of the distribution functions.
In macroscopic approaches  \cite{ay,den,dtlr} the isoscalar 
and isovector modes correspond to the in-phase and out-of-phase motions 
of neutrons and protons respectively. That means that both modes
can be described in terms of the distortions of the distribution
function in the form $\delta f^{(\pm)} = (\delta f_p \pm \delta f_n)/2$, 
where the subindices $p$ or $n$ label protons or neutrons and the plus 
or minus sign denotes the isoscalar or isovector modes, respectively.

We use in  the  nuclear interior the linearized the Vlasov- Landau 
equation for the dynamical component of the one-particle phase space 
distribution function $\delta f({\bf r},{\bf p},t)$, completed by 
a source term $J(\{f\})$ for relaxation processes. Neglecting a 
difference in the chemical potentials for protons and neutrons
and assuming $f_{0,p} = f_{0,n} = f_0$, where $f_0 \equiv 
f_0 ({\bf r}, {\bf p})$ is the equilibrium distribution function,
we write down the linearized two-component Vlasov- Landau equation 
in the form (symmetric nuclear matter approximation)
\begin{equation}   
{\partial \over \partial t} \delta f^{(\pm)} + {\vec{p}\over m^{*}}
\nabla_{r}\delta f^{(\pm)} - \nabla_{r}\delta V^{(\pm)} \nabla_{p}f_0 = 
J^{(\pm)}(\{\delta f^{(\pm)}\}), \ \ \ r<R_{0},            
\label{w1}
\end{equation}   
where $\delta V^{(\pm)} \equiv \delta V^{(\pm)} ({\bf r}, {\bf p}, t)$ is 
the Wigner transform of the variation of the self-consistent potential 
with respect to the equilibrium value $V_0$.  
This mean field variation can be expressed in terms of 
the interaction amplitude ${\cal F}^{(\pm)}({\bf p}, {\bf p}\prime)$ 
as  
\begin{equation}   
\delta V^{(\pm)} = {2 \over N_F} \int {d{\bf p}\prime 
\over (2 \pi {\hbar})^3}\,\, 
{\cal F}^{(\pm)}({\bf p}, {\bf p}\prime) \,\,
\delta f^{(\pm)}({\bf r}, {\bf p}\prime; t),
\label{w2}
\end{equation}   
where $N_F = p_F \,m^* /(\pi^2 \,\hbar^3), \,\,\,p_F$ is the Fermi 
momentum, $m^*$ is the effective mass of nucleon. The quantity 
${\cal F}^{(\pm)}({\bf p}, {\bf p}\prime)$ are defined by the interaction 
amplitudes $F_{\alpha,\beta}({\bf p}, {\bf p}\prime)$ 
between neutrons and protons $(\alpha,\beta) =(n, p)$:
$$
{\cal F}^{(\pm)}({\bf p}, {\bf p}\prime) =
(F_{p,p}({\bf p}, {\bf p}\prime) +F_{n,n}({\bf p}, {\bf p}\prime) 
+F_{p,n}({\bf p}, {\bf p}\prime)+ F_{n,p}({\bf p}, {\bf p}\prime))/2, 
$$
$$
{\cal F}^{(\pm)}({\bf p}, {\bf p}\prime) =
(F_{p,p}({\bf p}, {\bf p}\prime) +F_{n,n}({\bf p}, {\bf p}\prime) 
-F_{p,n}({\bf p}, {\bf p}\prime)- F_{n,p}({\bf p}, {\bf p}\prime))/2, 
$$
The amplitudes $F_{\alpha,\beta}({\bf p}, {\bf p}\prime)$ 
is usually parameterized in terms of the Landau constants 
$F_{\alpha,\beta,0}$ and $ F_{\alpha,\beta,1}$ as   
$
F_{\alpha,\beta}({\bf p}, {\bf p}\prime) =  
F_{\alpha,\beta, 0} + F_{\alpha,\beta, 1} ({\hat p} 
\cdot {\hat {p^\prime}}), \ \ \ {\hat p} = {\bf p}/p. 
$
This leads to the relation
\begin{equation}   
{\cal F}^{(\pm)}({\bf p}, {\bf p}\prime) =  
{\cal F}^{(\pm)}_0 + {\cal F}^{(\pm)}_1 ({\hat p} 
\cdot {\hat {p^\prime}}). 
\label{w3}
\end{equation}   
To simplify the presentation, we will omit in the following
the superscripts $(\pm)$ and include them only when it is 
necessary to avoid confusion.

  The right-hand side of Eq. (\ref{w1}) represents the change of the 
distribution function due to relaxation. We take into account the 
collisional damping exactly. The one-body relaxation is taken into 
consideration approximately by adding to the Vlasov-Landau equation 
some source terms. Namely, we assume
\begin{equation}                                 
J(\{\delta f\})
= J_c(\{\delta f\}) + J_s(\{\delta f\}),
\label{w4}
\end{equation}   
where $J_c(\{\delta f\})$ is the collision integral for the two-body 
collisions, $J_s(\{\delta f\})$ determines the change in the distribution 
function resulting from one-body relaxation.
The term $J_s(\{\delta f\})$ is considered within the 
relaxation time approximation  of the form
\begin{equation}   
J_s(\{\delta f\}) = -{ \delta f_0 ({\bf r},{\bf p},t) \over {\tau_{s,0}}}  
-{ \delta f_2 ({\bf r},{\bf p},t) \over {\tau_{s,2}}}.
\label{w5}
\end{equation}   
Here, $\delta f_{\ell }({\bf r},{\bf p},t)$ is dynamical component 
of the distribution function at the Fermi-surface distortion with 
multipolarity $\ell$ :
$   
\delta f({\bf r},{\bf p},t) =\sum_{\ell \geq 0} 
\delta f_{\ell }({\bf r},{\bf p},t)\equiv \sum_{\ell \geq 0} 
\delta \tilde{f}_{\ell }({\bf r},p,t) Y_{\ell 0}(\hat{p}),
$
and $\tau_{s,0},\tau_{s,1}$ is the relaxation time corresponding 
to the equilibration of the system due to the one-body dissipation.

The first component of the one-body source term $J_s(\{\delta f\})$ 
in Eq. (\ref{w5}) leads to nonconversation of energy. It determines 
an irreversible part of the energy  transferred from particles to the 
nuclear surface \cite{ts}.

The distribution functions in  the  nuclear interior are constructed 
as a linear  angular superposition  of  the corresponding
solutions ($ \delta f_{\vec{k}}(\vec{r},\vec{p},t)$) 
of the Vlasov- Landau  equation in nuclear matter:
$$
\delta f_{\lambda,k}(\vec{r},\vec{p},t) = Re \int d\Omega _{k}Y_{\lambda 0}
(\hat{k})\delta f_{\vec{k}}(\vec{r},\vec{p},t),
$$
\noindent where
\begin{equation}   
\delta f_{\vec{k}}(\vec{r},\vec{p},t) 
= - {\partial f^{eq} \over \partial \epsilon^{eq}}
\exp {\{i(\vec{k}\vec{r}-\omega t)\}}
\sum^{}_{\ell \geq 0} \alpha_{\ell }(\omega,k)
Y_{\ell 0}(\hat{p} \cdot \hat{k}). 
\label{w6}
\end{equation}   

 Substituting these expressions into the Vlasov- Landau equation, 
integrated with respect to the energy $\epsilon_1$, we obtain an
equation for $\alpha_{\ell}$  and the velocity 
$S$ = $\omega / (v_{F}k)$. In order for the closed-form results to
obtain, we will follow the nuclear fluid dynamic approach of Refs. 
\cite{he}-\cite{na} and take into account in Eq.(\ref{w1}) the dynamic 
Fermi-surface distortions up to multipolarity $\ell = 2$.
As a result we have the system 
\begin{eqnarray}
\alpha_{0} - (1+{\cal F}_{1}/ 3)\alpha_{1}/ \sqrt {3}  
&=& -  i / ( v_{F} k \bar{\tau}_{0})\alpha_{2}\,, 
\nonumber
\\ 
\alpha_{1} - (1+{\cal F}_{0})\alpha_{0}/ \sqrt {3} -
 \alpha_{2} /\sqrt {15} 
&=&  -  i / ( v_{F} k \bar{\tau}_{1})\alpha_{2}\,,
\label {w7}
\\
\alpha_{2} - 2(1+{\cal F}_{1}/ 3)\alpha_{1}/ \sqrt {15}
&=&  -  i / ( v_{F} k \bar{\tau}_{2})\alpha_{2}\,.
\nonumber
\end{eqnarray} 

  The effective relaxation times $\bar{\tau}_{\ell}$ are different for 
isoscalar and isovector modes: 
\begin{equation}
\bar{\tau}^{(\pm)}_{0} = \tau^{(\pm)}_{s,0}\,, 
\bar{\tau}^{(-)}_{1} = \tau^{(-)}_{1}\,, 1/ \bar{\tau}^{(+)}_{1} = 0\,, 
1/\bar{\tau}^{(\pm)}_{2} = 1/\tau^{(\pm)}_{2} + 1/ \tau^{(\pm)}_{s,2}\,.
\label {w8}
\end{equation} 
Here $\tau^{(\pm)}_{\ell}$ are the partial collective relaxation times 
due to interparticle collisions within the distorted layers of the
Fermi surface with multipolarity $\ell$. These times enter in the multipole 
expansions of the total numbers ${\cal N^{(\pm)}}(\hat{p})$ of the 
collisions in direction $\hat{p} \equiv \hat{p}_1$:

\begin{equation}
{\cal N^{(\pm)}}(\hat{p}) \equiv
\int^{\infty}_{0} d\epsilon_{1} J^{(\pm)}_c(\hat{p},\epsilon_{1})
= \exp{\{-i\omega t\}}\sum^{}_{\ell \ge \ell^{(\pm)}_{0}}
\sum^{\ell }_{m=-\ell }
{\alpha^{(\pm)}_{l m}\over \tau^{(\pm)}_{\ell}} 
Y_{\ell m}(\hat{p}) ,
\label{w9}
\end{equation}
\noindent where $\ell^{(-)}_{0} =1$, $\ell^{(+)}_{0} =2$ ,
\begin{equation}      
{1\over \tau^{(\pm)}_{\ell }} \equiv  \int^{\infty }_{0}
d\epsilon_1 \int d\Omega_p J^{(\pm)}_c(\hat{p},\epsilon_{1})
Y_{\ell 0}(\hat{p})/
\int^{\infty }_{0}d\epsilon_{1} \int d\Omega_{p}
\delta f^{(\pm)} Y_{\ell 0}(\hat{p}).
\label{w10}
\end{equation}

 With the collision integrals with memory effects \cite{kmp}- \cite{kp1} 
for interaction between different kinds of particles and following the 
procedure \cite {akh} in integrating the collision integral over energy, 
we obtain the following expressions for relaxation times at 
$\hbar \omega ,T \ll E_F$:

$$
1 / \tau^{(+)}_{\ell }   
= {\cal R}(\omega, T)<({\bar W}+W_{pn}) \Phi^{(+)}_{\ell }>,
$$
\begin{equation}
1 / \tau^{(-)}_{\ell }   
= {\cal R}(\omega, T)<{\bar W} \Phi^{(+)}_{\ell }>+
<W_{pn} \Phi^{(-)}_{\ell }>,
\label{w11}
\end{equation}
where ${\bar W}=(W_{nn}+W_{pp})/2$; $W_{\alpha,\beta}$ is the probability
of scattering of the particles $(\alpha,\beta) =(n, p)$ near the Fermi 
surface. The functions 
$
\Phi^{(+)} _{\ell } \equiv
1 + P_{\ell }({\hat{p}}_{2}{\hat{p}}_{1}) -
P_{\ell }({\hat{p}}_{3}{\hat{p}}_{1}) -
P_{\ell }({\hat{p}}_{4}{\hat{p}}_{1})$ ,
$
\Phi^{(-)} _{\ell } \equiv
1 - P_{\ell }({\hat{p}}_{2}{\hat{p}}_{1}) -
P_{\ell }({\hat{p}}_{3}{\hat{p}}_{1}) +
P_{\ell }({\hat{p}}_{4}{\hat{p}}_{1}) ,
$
define the angular constraints for nucleon's scattering 
($P_{\ell }$ is a Legendre polynomial;${\vec p}_1$, ${\vec p}_2$ 
and ${\vec p}_3$, ${\vec p}_4$ are the momentum of particles before and 
after collisions respectively). The symbol $<\ldots >$ denotes the 
averaging over angles of the relative momentum of the colliding 
particles. The function ${\cal R}(\omega, T)$ has the following form
\begin{equation}
{\cal R}(\omega, T)
= \left( {m^{*}\over \hbar ^{2}} \right)^{3} {1\over 48\pi ^{4}} 
\{(2\pi T)^{2}+ C_{\omega }(\hbar \omega)^{2}\},
\label{w12}
\end{equation}
where we will take $C_{\omega } =1$ \cite {kps1}. We have the following 
relations in the case of isotropic collision probabilities
$$
1 / \tau^{(-)}_{1} = (4/3) {\cal R} <W_{pn}> = 
(5 W_{pn}/3({\bar W} + W_{pn}))/ \tau^{(+)}_{2} 
\approx 1.1 /\tau^{(+)}_{2} ,
$$
\begin{equation} 
\tau^{(-)}_{2} = \tau^{(+)}_{2}.
\label {w13}
\end{equation}

We find the following  dispersion equation for calculation 
the velocity $S$ from Eq.(\ref{w7})    
\begin{equation}   
\left ( S+ { i \over v_{F} k \tau_{0}} \right) 
\left ( S+ { i \over v_{F} k \tau_{1}} 
- \left({4\over 15}\right)(1+ {\cal F}_1/3) / 
\left({S + {i\over v_{F}k \tau_{2}}}\right) \right] -
S^{2}_{f}=0,
\label{w14}
\end{equation}   
where $S_{f}$ is the velocity of the first sound: 
$
S^{2}_{f}= (1+{\cal F}_0)(1+ {\cal F}_1/3)/3 .
$

 The complex frequency  of the collective vibrations of the
$\lambda $ type is determined by the velocity $S = S_{\lambda}$  
and the wave number $k_{\lambda}$:  
$\omega_{\lambda} =  v_{F}k_{\lambda}S_{\lambda}$.
The values $k_{\lambda, n}$ of the wave number $k_{\lambda}$ can be found 
from the macroscopic boundary conditions acting at the nuclear surface 
\cite{kmp,den}. 
Here we study the damping properties of giant resonances rather than 
their full description and parametrize damping width as a function of the
resonance energy. For underdamped collective motion the quantity
$k_{\lambda, n}$ can be considered as a real number. Then according to the 
correspondence principle the energies ${\cal E}_{\lambda,n}$ and widths
$\Gamma_{\lambda,n}$ of the resonances are given by
\begin{eqnarray}
{\cal E}_{\lambda,n} &=& \hbar \Omega_{\lambda,n} = 
\hbar v_{F}k_{\lambda,n}S^{(r)}_{\lambda},
\nonumber
\\
\Gamma_{\lambda,n} &=& -2\hbar Im \{\omega_{\lambda,n}\} = 
2 \hbar v_{F}k_{\lambda,n}S^{(i)}_{\lambda},
\label{w15}
\end{eqnarray} 
\noindent where $\Omega \equiv Re \omega$,
$S^{(r)} \equiv Re S$, $S^{(i)} \equiv - Im S$.

We consider the case of the slightly damped motion when 
condition $\mid~S_i~/~S_0~\mid~\ll~a$ is fulfilled, where
$a = min \{1, x, 1/x, 1/(1/y+ 1/z) \}$ and $ x \equiv \Omega \tau_{2}$,
$ y \equiv \Omega \tau_{0}$, $ z \equiv \Omega \tau_{1}$.
As a first approximation, we obtain from Eq.(\ref{w14})

\begin{eqnarray}
2{S^{(i)} \over S^{(r)}} &=& 
{\bar {S}^{2}_{0} - \bar {S}^{2}_{f} \over (S^{(r)})^{2}}
{x \over 1+x^{2}} \left( 1- {x \over y} \right) 
\left( 1- {1 \over y z} \right) + { 1 \over y} + { 1 \over z},
\label{w16}
\\
\left( S^{(r)} \right)^{2} &=&  \bar{S}^{2}_{0} +
\left( \bar{S}^{2}_{f} - \bar{S}^{2}_{0} \right) {1 \over 1+ x^{2}}
\left( 1 - { x \over y} \right),
\label{w17}
\end{eqnarray} 
\noindent where $\bar{S}^{2}_{f} = S^{2}_{f}/( 1 - 1/(yz))$, 
$\bar{S}^{2}_{0} = S^{2}_{0}/( 1 - 1/(yz))$ and
$S^{2}_{0} = S^{2}_{f}+ (4/15)(1+ {\cal F}_1/3)$ .
The quantity  $S_{0}$ is the velocity of the zero sound in 
the absence of the relaxation processes and with the deformation 
of the Fermi sphere with multipolarities $ \ell \leq 2$ only.

 We find the following expressions for the intrinsic width and
frequency $\Omega = {\cal E} / \hbar $  of a GMR using Eqs. 
(\ref{w15}- \ref{w17}):  
\begin{eqnarray}
\Gamma &=& \hbar\Omega (a-1){x \over 1+ a( x^2 +x/y)/(1-x/y)}
(1-{1 \over yz}) + {\hbar \Omega \over y}+ {\hbar \Omega \over z},
\label{w18}
\\
\Omega^{2} &=&  \omega^{2}_{0} +
\left( \omega^{2}_{f} - \omega^{2}_{0} \right) {1 \over 1+ x^{2}}
\left( 1 - { x \over y} \right),
\label{w19}
\end{eqnarray}
\noindent where $a=(E_{0}/ E_{f})^2$ and 
$E_{0} = \hbar \omega_{0} = \hbar v_{F}k \bar{S}^{(r)}$ , 
$E_{f} = \hbar \omega_{f} = \hbar v_{F} k \bar{S}^{(f)}$. 

Note that the total change rate in the distribution function was 
taken as a sum of the change rates in various damping channels 
(independent dissipation rates approximation) but 
in the general case the expression for $\Gamma$ can not 
be represented as a sum of the widths associated with the different 
independent sources of the damping. This is a peculiarity of the 
collisional Vlasov- Landau equation where the Fermi- surface 
distortion effect influences both the self- consistent mean field and 
the memory effect at the relaxation processes.

\section{ The numerical results and discussion}

The values of the GMR energy and the relaxation times  
are required for calculations of the intrinsic 
width $\Gamma $. As the GMR energy ${\cal E}$  we use the 
phenomenological $A$-dependence of ${\cal E}$ obtained from a 
fit to the experimental data \cite{bf} -\cite{woude}. 
We neglect of the variation of the wave number $k$ in the first 
and zero sound regimes ( \cite{ehp,na}) and adopt  for energy
$E_{f}$  the values corresponding to the energy in
hydrodynamic approach. We use also the approximation 
$ E_{0}\simeq {\cal E}$ due to consideration of the underdapmed motion.  
The estimation from \cite{kps} is used for the collisional 
relaxation time $\tau^{(+)}_{2}$ and for one-body relaxation time
$\tau^{(-)}_{s,0}$ ($\xi^{1} = 1.543$; $\tau^{(-)}_{s,2} = 0$). 

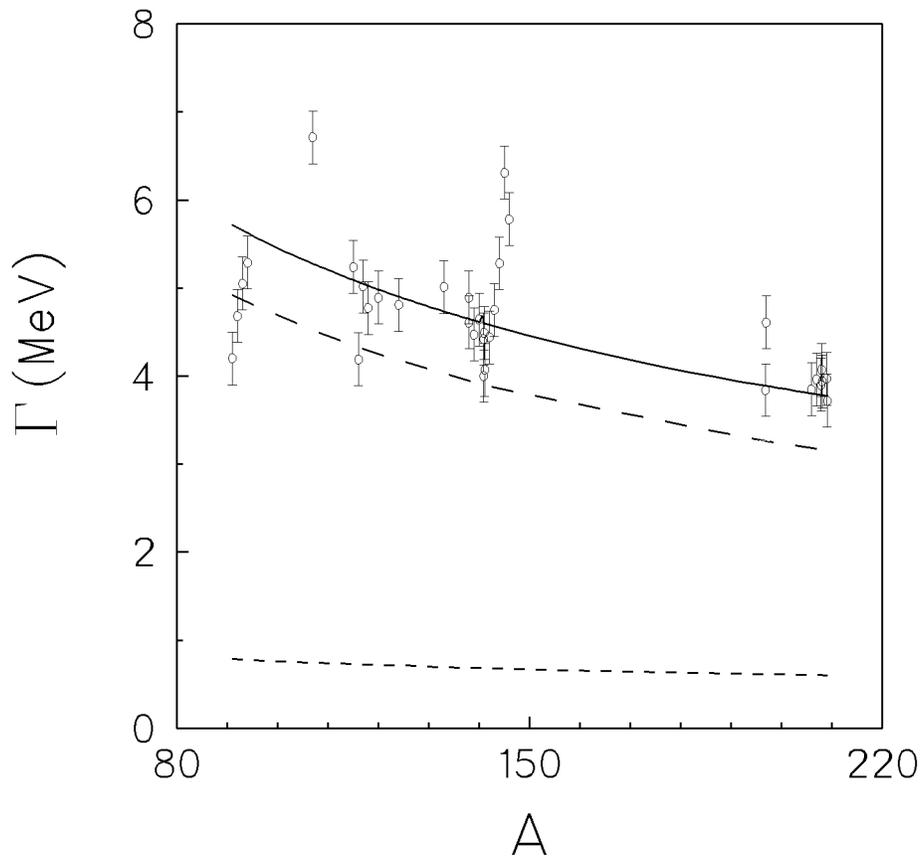
In Fig. 1 we show the intrinsic  the giant dipole resonance (GDR) 
widths and the one- and the two- body contributions to them at 
zero temperature ($T=0$) as functions of mass number.
The experimental data were taken from \cite{bf}.  
The contribution of collisional damping (dot-dash line with 
long dash) to the GDR widths does not exceed $\sim 30\%$ of the 
experimental values. 

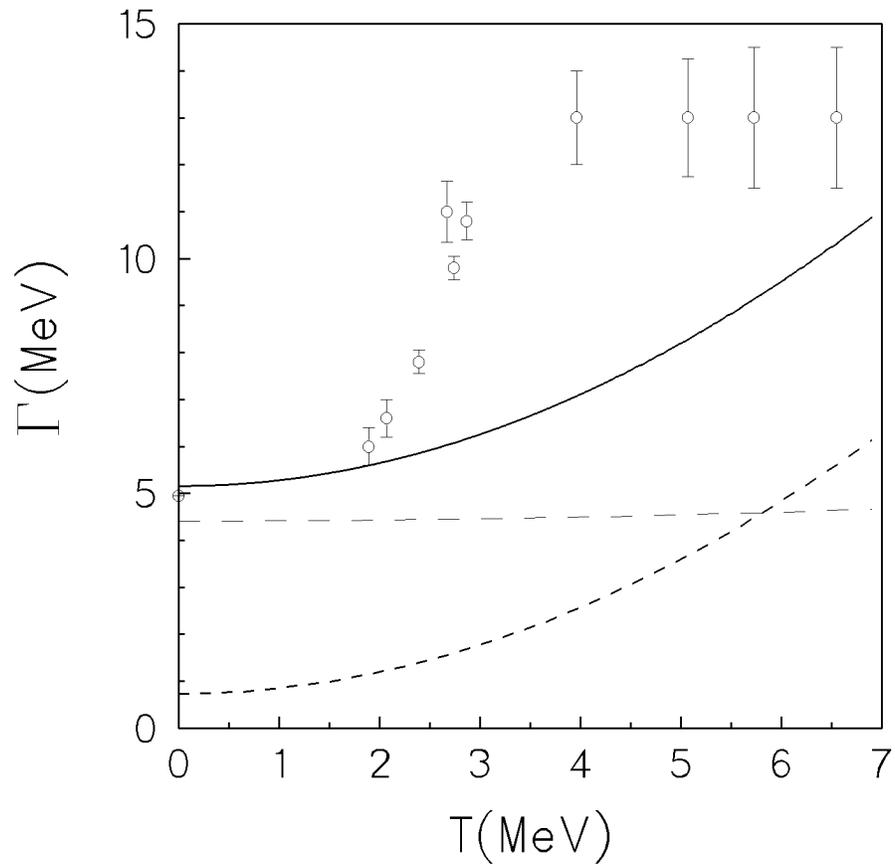
In Fig. 2 we show the intrinsic width of the giant dipole resonance  
(GDR) in the nucleus $^{112}${\it Sn} as a function of the temperature
$T$. The experimental data were taken from \cite{g92}. Considering the 
experimental data we assumed that the energy ${\cal E}$ of the GDR is 
independent of temperature and equals $15.6 MeV$. Note that we can use 
the expression (\ref{w18}) for intrinstic width  when 
the condition  $\Gamma / {\cal E} \ll 1$  is fulfilled, i.e., in fact, 
at the temperature of no more than $\approx 5 MeV$. We observe a systematic 
large deviation of the evaluated width with respect to the experimental data 
in the range from 2 to 5 Mev. This deviation may be connected with 
dependence of the GDR width on angular momentum and thermodynamic 
fluctuations  of the nuclear  shape  and orientation angles \cite{mbc}.\\

\vspace{0.7cm}

This  work  was supported in part by the IAEA under Contract  No. 10308/RBF. 


\bigskip


\end{document}